\documentclass[onecolumn,secnumarabic,amssymb, nobibnotes, aps, pre,groupedaddress,notitlepage]{revtex4-1}

\usepackage{graphicx}
\usepackage{xcolor}
\usepackage{amsmath}
\usepackage{float}
\usepackage{soul}
\begin{document}

\title{Generalized Bloch's theorem for viscous metamaterials: Dispersion and effective properties based on frequencies and wavenumbers that are simultaneously complex}

\author{Michael J. Frazier}
\thanks{Current affiliation: California Institute of Technology}
\author{Mahmoud I. Hussein}
\email[Email: ]{mih@colorado.edu}
\affiliation{Department of Aerospace Engineering Sciences, University of Colorado Boulder, Boulder, CO 80309-0429, USA}

\date{\today}

\begin{abstract}
It is common for dispersion curves of damped periodic materials to be based on real frequencies versus complex wavenumbers or, conversely, real wavenumbers versus complex frequencies. The former condition corresponds to harmonic wave motion where a driving frequency is prescribed and where attenuation due to dissipation takes place only in space alongside spatial attenuation due to Bragg scattering. The latter condition, on the other hand, relates to free wave motion admitting attenuation due to energy loss only in time while spatial attenuation due to Bragg scattering also takes place. Here, we develop an algorithm for 1D systems that provides dispersion curves for damped free wave motion based on frequencies and wavenumbers that are permitted to be simultaneously complex. This represents a generalized application of Bloch's theorem and produces a dispersion band structure that fully describes all attenuation mechanisms, in space and in time. The algorithm is applied to a viscously damped mass-in-mass metamaterial exhibiting local resonance. A frequency-dependent effective mass for this damped infinite chain is also obtained.    
\end{abstract}

\maketitle
%\tableofcontents

\section{Introduction}
\label{sec:introd}
Phononic materials have been attracting much attention in the materials physics and engineering communities because of their rich scope of acoustic and elastodynamic properties \cite{MeadJSV1996,HusseinAMR2014,Deymier2013,Khelif2015,Laude2015}. There are two classes of phononic materials: phononic crystals \cite{SigalasJSV1992,KushwahaPRL1993} and locally resonant acoustic/elastic metamaterials \cite{LiuScience2000}. Phononic crystals resemble atomic-scale crystals in that they consist of repeated units in space. The dispersion curves for elastic wave propagation in a phononic crystal, or a periodic material in general, appear in bands and in some cases band gaps may arise. Bragg scattering is the prime band-gap opening mechanism in a phononic crystal. Metamaterials are usually also periodic, although not by necessity. The prefix \it{meta-}\rm ~(meaning ``beyond") is associated with the notion that through deliberate design of the internal structure, these materials manifest unusual properties that exceed those of conventional composites and phononic crystals. In the context of wave propagation, local resonances are generally the key feature in metamaterials leading to salient properties such as subwavelength band gaps \cite{LiuPRB2002,WangPRL2004,PennecPRB2008,WuAPL2008}, negative effective material properties \cite{LiPRE2004,DingPRL2007,AoWRCM2010,LiuAPL2011}, enhanced dissipation \cite{HusseinJSV2013,AntoniadisJSV2015,Chen_Sun_2016}, and thermal conductivity reduction \cite{DavisHusseinPRL2014}, among others.\\
\indent At present, much of the phenomena of wave propagation in phononic materials is understood from the perspective of conservative linear elasticity. Realizing their full potential, however, requires an account of energy dissipation from damping. Already, metamaterials possessing internal resonating bodies have been shown to demonstrate enhanced dissipation under certain conditions (i.e., beyond which may be attributed to the sum of the individual material constituents) \cite{HusseinJSV2013,AntoniadisJSV2015,Chen_Sun_2016}. This property is beneficial where enhanced dissipation in a structure is needed but without appearing at the expense of stiffness. There are numerous avenues for the treatment of damping in material or structural models, including those representing phononic materials. A common approach is to consider viscous damping, for which a simple version is known as Rayleigh \cite{Rayleigh1877}, or proportional, damping$-$whereby the matrix of damping coefficients is assumed to be proportional to the mass and/or stiffness matrices \cite{CaugheyJAM1965,AdhikariJSV2006,AdhikariJVA2009}. If the proportionality condition is not met, then the model is described as \it{generally damped}\rm  ~\cite{WoodhouseJSV1998,AdhikariJSV2001}. Experiments are used to determine an appropriate damping model for a given material or structure \cite{PhaniJSV2007}.\\
\indent Beyond the choice of the damping model, an important consideration is whether the frequency or the wavenumber is selected to be real and, consequently, which is permitted to be complex. There are two classes of problems dealing with damped phononic materials. In one class, the frequencies are assumed \it{a priori}\rm ~to be real thus allowing the effects of damping to manifest only in the form of complex wavenumbers. Physically, this represents a medium experiencing wave propagation due to a sustained driving frequency and dissipation taking effect in the form of spatial attenuation only. This approach follows a $\kappa=\kappa(\omega)$ formulation (where $\kappa$ and $\omega$ denote wavenumber and frequency, respectively) resulting from either a linear \cite{TassillyIJES1987,LangleyJSV1994,LaudePRB2009,RomeroNJP2010,MoiseyenkoPRB2011,AndreassenJVA2013} or a quadratic \cite{MeadJSV1973,FarzbodJVA2011,ColletIJSS2011} eigenvalue problem (EVP). In the other class, the frequencies are permitted to be complex thus allowing dissipation to take effect in the form of temporal attenuation. Physically, this represents a medium admitting free dissipative wave motion, e.g., due to impulse loading. Here, a $\omega=\omega(\kappa)$ formulation leading to a linear EVP is the common route (in some cases with the aid of a state-space transformation); see Refs.~\cite{MukherjeeCS1975,SprikSSC1998,HusseinPRB2009,HusseinJAP2010,PhaniJVA2013,AndreassenJVA2013}. \\
\indent In the `driven waves' path, a real frequency is prescribed and the underlying EVP is solved for a corresponding pair of real and imaginary wavenumbers, representing propagation and attenuation constants, respectively. All modes are described by complex wavenumbers due to the dissipation. In the `free waves' path, on the other hand, a real wavenumber is specified, and complex frequencies emerge as the solution (the real and imaginary parts respectively provide the loss factor and the frequency for each mode). Because of the common association of the driven waves problem to an EVP for which the frequency is the independent variable and, in contrast, the free waves problem to an EVP for which the wavenumber is the independent variable, it is often viewed that the two only available options are: real frequencies and complex wavenumbers versus real wavenumbers and complex frequencies~\cite{Achenbach1999,MaceJSV2008,ManconiJSV2010}. However, if the medium permits spatial attenuation in its undamped state$-$which is the case for phononic materials within band-gap frequencies$-$then, in principle, there should be an imaginary wavenumber component (in addition to the real wavenumber component) even when the frequencies are complex. This, in fact, represents a more complete picture of the dispersion curves for damped free wave motion in media that contain inherent mechanisms for spatial attenuation, such as Bragg scattering and local resonance. Since this scenario pertains only to free waves, one expects to see complex frequencies for bands admitting only spatial propagation as well as bands admitting evanescent waves (with the real part of the wavenumber being either zero or $\pi$ divided by the lattice spacing$-$the two values that represent the limits of the irreducible Brillouin zone). For a proportionally damped problem, a solution that permits both the frequencies and wavenumbers to be complex has been obtained using the transfer matrix method which gives a $\kappa=\kappa(\omega)$ linear EVP~\cite{HusseinHB2013}. For a generally damped problem, however, an all-complex %(i.e., both frequencies and wavenumbers being gnerally complex) 
solution cannot be obtained from a linear EVP, nor from directly solving a quadratic EVP.\\
\indent In this paper, we consider damped free motion in 1D systems and provide an algorithm$-$based on a quadratic EVP$-$that provides the dispersion curves and damping ratio constants for both spatially propagating and attenuating waves. As an example, we focus on a viscously damped mass-in-mass chain representing a locally resonant acoustic/elastic metamaterial~\cite{HuangIJES2009}. With a complex wavenumbers-complex frequencies band structure at hand, we also compute a frequency-dependent effective mass. In the absence of dissipation, this effective mass is real and negative in the region of the band gap. As damping alters the dispersion characteristics, we observe the effective mass to transition to complex form and its region of negativity diminishes as the intensity of the damping increases.\\

\section{Theory: Dispersion Relations}
\begin{figure}[h]
	\centering
	\includegraphics[scale= 0.8]{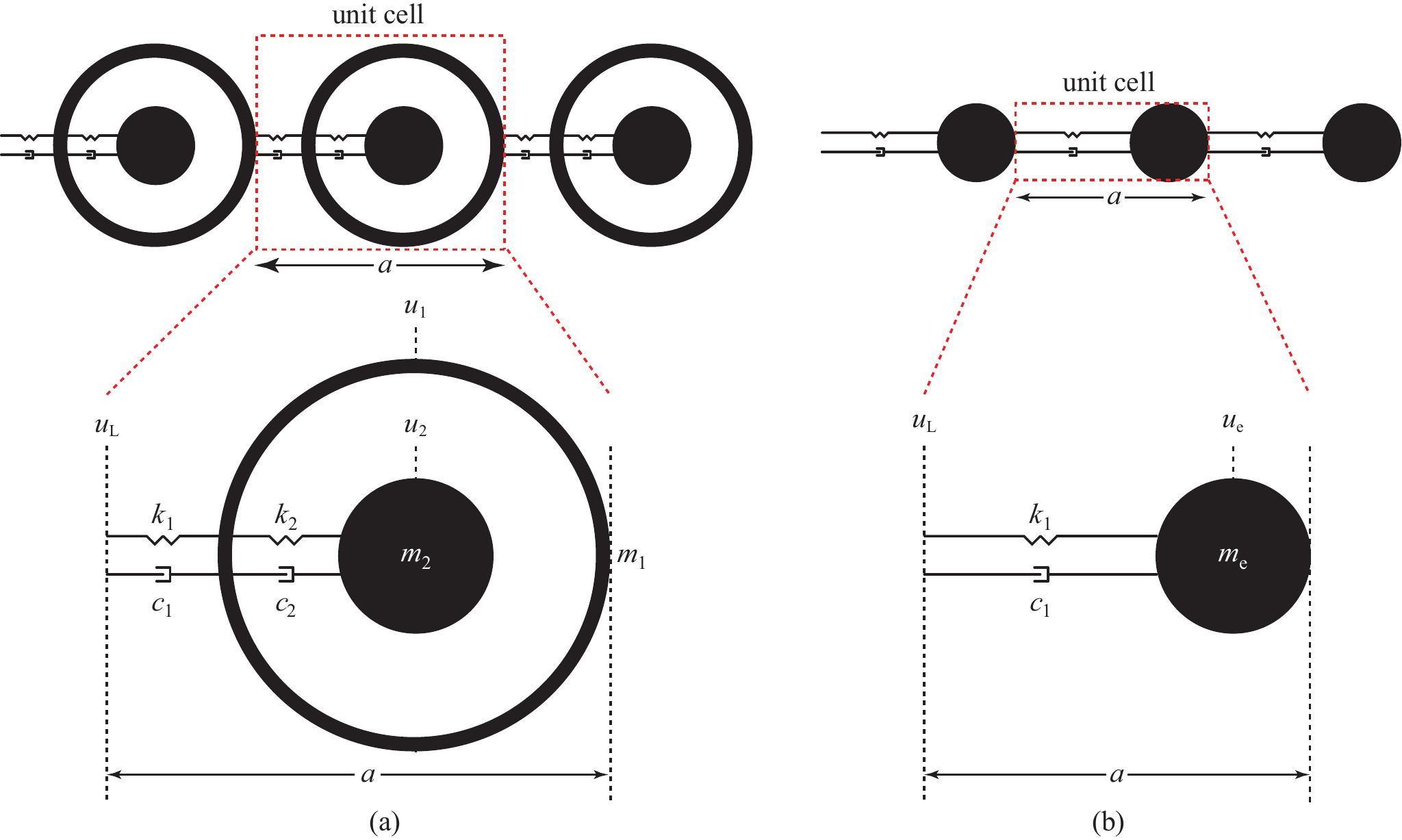}
	\caption{One-dimensional, discrete unit cell for a periodic material with lattice spacing $a$: (a) 2-DOF acoustic/elastic metamaterial; (b) equivalent 1-DOF material}
	\label{fig:FIG_01_Model}
\end{figure}

We consider a nested 1D lumped parameter mass-spring-dashpot model similar to what is investigated in Ref.~\cite{HuangIJES2009}. Infinite in extent, a model of a locally resonant acoustic/elastic metamaterial is constructed by appending copies \it{ad infinitum}\rm~of the unit cell depicted in Fig.~\ref{fig:FIG_01_Model}a along the line of motion. Prior to application of boundary conditions, there are three degrees of freedom (DOF) in this unit cell with $u_\mathrm{L}$, $u_1$, and $u_2$ denoting the displacement of the left and right cell boundaries and an internal DOF, respectively. Each of these have an associated mass: $m_\mathrm{L}=0$, $m_1$, and $m_2$. On the boundaries, the forces $f_\mathrm{L}$ and $f_1$ apply. Springs with stiffness $k$ and dashpots with viscosity $c$ connect the DOFs.

%In this work, we apply the Bloch boundary condition to the unit cell to formulate a quadratic eigenvalue problem whose solution, given an input frequency, is a complex wavenumber, $\kappa$. With material damping, we assume the input frequency, $\lambda$, to be complex, representing wave propagation with temporal decay, a characteristic of free waves. Prescribed waves do not attenuate in time due to damping and $\lambda=\mathrm{i}\omega$ is always true. Thus, two band diagrams emerge from the dispersion relations: one accounting for the wave frequency and a second quantifying the wave temporal decay.

Balancing all dynamic forces, the motion of each DOF in Fig.~\ref{fig:FIG_01_Model}a is described by the following equations:
\begin{subequations}
	\label{eq:eom_nodal}
	\begin{align}
	&m_1\ddot{u}_1+(c_1+c_2)\dot{u}_1-c_2\dot{u}_2-c_1\dot{u}_\mathrm{L}+(k_1+k_2)u_1-k_2u_2-k_1u_\mathrm{L}=f_1,\\
	&m_2\ddot{u}_2-c_2(\dot{u}_1-\dot{u}_2)-k_2(u_1-u_2)=0,\\
	&m_\mathrm{L}\ddot{u}_\mathrm{L}-c_1(\dot{u}_1-\dot{u}_\mathrm{L})-k_1(u_1-u_\mathrm{L})=f_\mathrm{L}.
	\end{align}
\end{subequations}
%where $u_\mathrm{L}$ is the displacement of the left \textcolor{red}{boundary} of the unit cell, $f_\mathrm{L}$ is the force acting on this \textcolor{red}{boundary}, and $m_\mathrm{L}=0$ is the associated mass.
Together, Eqs. \eqref{eq:eom_nodal} may be assembled into a system of equations
\begin{equation}
\label{eq:eom_mat_1}
\mathbf{M}\ddot{\mathbf{u}}+\mathbf{C}\dot{\mathbf{u}}+\mathbf{K}\mathbf{u}=\mathbf{f},
\end{equation}
where $\mathbf{M}$, $\mathbf{C}$, and $\mathbf{K}$ are the assembled mass, damping, and stiffness matrices, respectively. Collecting and arranging the nodal displacements $\mathbf{u}^\mathrm{T}=[u_1\;\;u_2\;\;u_\mathrm{L}]$ and forces $\mathbf{f}^\mathrm{T}=[f_1\;\;0\;\;f_\mathrm{L}]$, allows the mass, damping, and stiffness matrices to be defined as follows:
\begin{subequations}
	\begin{align}
	&\mathbf{M}=m_2\mathbf{M}_\mathrm{r}=m_2
	\begin{bmatrix}
	1/r_\mathrm{m} & 0 & 0\\
	0 & 1 & 0\\
	0 & 0 & 0
	\end{bmatrix},\\
	&\mathbf{C}=c_2\mathbf{C}_\mathrm{r}=c_2
	\begin{bmatrix}
	1/r_\mathrm{c}+1 & -1 & -1/r_\mathrm{c}\\
	-1 & 1 & 0\\
	-1/r_\mathrm{c} & 0 & 1/r_\mathrm{c}
	\end{bmatrix},\\
	&\mathbf{K}=k_2\mathbf{K}_\mathrm{r}=k_2
	\begin{bmatrix}
	1/r_\mathrm{k}+1 & -1 & -1/r_\mathrm{k}\\
	-1 & 1 & 0\\
	-1/r_\mathrm{k} & 0 & 1/r_\mathrm{k}
	\end{bmatrix},
	\end{align}
\end{subequations}
where we utilize the ratios $r_\mathrm{m}=m_2/m_1$, $r_\mathrm{c}=c_2/c_1$, and $r_\mathrm{k}=k_2/k_1$ to write the matrices in terms of the resonator parameters. 
%Following our earlier assumption, the displacement functions take the form $\mathbf{u}=\tilde{\mathbf{u}}\mathrm{e}^{\lambda{t}}$.
Dividing by $m_2$, the system in Eq. \eqref{eq:eom_mat_1} is written as
\begin{equation}
\label{eq:eom_mat_2}
\mathbf{M}_\mathrm{r}\ddot{\mathbf{u}}+\beta\mathbf{C}_\mathrm{r}\dot{\mathbf{u}}+\omega_0^2\mathbf{K}_\mathrm{r}\mathbf{u}=\mathbf{f}/m_2,
\end{equation}
where $\beta=c_2/m_2$ is a measure of damping intensity (which varies with $c_2$,) and $\omega_0^2=k_2/m_2$ is the resonance frequency.

In the periodic material, $u_\mathrm{L}$ is tied to $u_1$ of the previous unit cell; likewise, $u_1$ is tied to $u_\mathrm{L}$ of the subsequent unit cell. For the material to support Bloch wave propagation, the displacements at the boundaries are related according to $u_\mathrm{L}=\mathrm{e}^{-\mathrm{i}\kappa{a}}u_1$. If we define the essential set of displacements as $\bar{\mathbf{u}}^\mathrm{T}=[\mathbf{u}_\mathrm{b}\;\;\mathbf{u}_\mathrm{i}]$ ($\mathbf{u}_\mathrm{b}$ gathers the essential boundary displacements and $\mathbf{u}_\mathrm{i}$ collects all the internal displacements), then the complete set of displacements may be written in terms of $\bar{\mathbf{u}}$ and the translational operator $\mathbf{T}$ as follows:
\begin{equation}
\label{eq:bloch_boundaries}
\mathbf{u}=\mathbf{T}\bar{\mathbf{u}}. 
\end{equation}
Given $\mathbf{u}_\mathrm{b}=u_1$ and $\mathbf{u}_\mathrm{i}=u_2$,
\begin{equation}
\mathbf{T}=
\begin{bmatrix}
1 & 0\\
0 & 1\\
\gamma & 0
\end{bmatrix}, \hspace{2mm} \gamma=\mathrm{e}^{-\mathrm{i}\kappa{a}}.
\end{equation}
The Bloch boundary condition is applied via the matrix $\mathbf{T}$. % At this point, it is important to note that $\kappa$ is generally complex.
Substituting Eq. \eqref{eq:bloch_boundaries} into Eq. \eqref{eq:eom_mat_2} and premultiplying by $\mathbf{T}^*$ (the conjugate transpose of $\mathbf{T}$), we arrive at 
\begin{equation}
\label{eq:eom_mat_3}
\bar{\mathbf{M}}_\mathrm{r}\ddot{\bar{\mathbf{u}}}+\beta\bar{\mathbf{C}}_\mathrm{r}\dot{\bar{\mathbf{u}}}+\omega_0^2\bar{\mathbf{K}}_\mathrm{r}\bar{\mathbf{u}}=\mathbf{0},
\end{equation}
where
\begin{subequations}
	\label{eq:mat_def}
	\begin{align}
	&\bar{\mathbf{M}}_\mathrm{r}=\mathbf{T}^*\mathbf{M}_\mathrm{r}\mathbf{T},\\
	&\bar{\mathbf{C}}_\mathrm{r}=\mathbf{T}^*\mathbf{C}_\mathrm{r}\mathbf{T},\\
	&\bar{\mathbf{K}}_\mathrm{r}=\mathbf{T}^*\mathbf{K}_\mathrm{r}\mathbf{T}.
	\end{align}
\end{subequations}
Equilibrium in the region between unit cells leads to $\mathbf{T}^*\mathbf{f}=\mathbf{0}$ \cite{FarzbodJVA2011propagation}. For general wave motion, a displacement solution takes the form $\bar{\mathbf{u}}=\tilde{\mathbf{u}}\mathrm{e}^{\lambda{t}}$. For driven waves, $\lambda=-\mathrm{i}\omega$, while for free waves, $\lambda$ is generally complex and yet to be determined. Applying this solution form to Eq. \eqref{eq:eom_mat_3} yields
\begin{equation}
\label{eq:eom_mat_4}
(\lambda^2\bar{\mathbf{M}}_\mathrm{r}+\lambda\beta\bar{\mathbf{C}}_\mathrm{r}+\omega_0^2\bar{\mathbf{K}}_\mathrm{r})\tilde{\mathbf{u}}=\mathbf{0},
\end{equation}
which, upon simplification, gives the following quadratic matrix relationship for a non-trivial solution:
\begin{equation}
\lambda^2\mathbf{A}+\lambda\mathbf{B}+\mathbf{I}=\mathbf{0},
\end{equation}
where $\mathbf{A}={1/\omega_0^2}\bar{\mathbf{K}}_\mathrm{r}^{-1}\bar{\mathbf{M}}_\mathrm{r}$ and $\mathbf{B}={\beta/\omega_0^2}\bar{\mathbf{K}}_\mathrm{r}^{-1}\bar{\mathbf{C}}_\mathrm{r}$. This formulation may proceed along two paths resulting in either generalized frequency solutions, $\lambda(\kappa)$, or generalized wavenumber solutions, $\kappa(\lambda)$.

\subsection{Frequency Solutions from Linear Eigenvalue Problem} \label{sec:solution_freq}
%When we implement the derivatives, the outcome is an eigenvalue problem of unconventional form:
%\begin{equation}
%	[\lambda^2\mathbf{M}+\lambda\mathbf{C}(\kappa)+\mathbf{K}(\kappa)]\mathbf{U}=\mathbf{0}.
%\end{equation}
In the case of a prescribed, real-valued wavenumber, the matrices in  Eq. \eqref{eq:eom_mat_3} are written explicitly as follows:
\begin{subequations}
	\label{eq:mat_def_freq}
	\begin{align}
	&\bar{\mathbf{M}}_\mathrm{r}=
	\begin{bmatrix}
	1/r_\mathrm{m} & 0\\
	0 & 1
	\end{bmatrix},\\
	&\bar{\mathbf{C}}_\mathrm{r}=
	\begin{bmatrix}
	2(1-\cos\kappa{a})/r_\mathrm{c}+1 & -1\\
	-1 & 1
	\end{bmatrix}, \label{eq:mat_def_freq_b}\\
	&\bar{\mathbf{K}}_\mathrm{r}=
	\begin{bmatrix}
	2(1-\cos\kappa{a})/r_\mathrm{k}+1 & -1\\
	-1 & 1
	\end{bmatrix}. \label{eq:mat_def_freq_c}
	\end{align}
\end{subequations}

In general, $\bar{\mathbf{C}}_\mathrm{r}$ is not simultaneously diagonalizable with $\bar{\mathbf{M}}_\mathrm{r}$ and/or $\bar{\mathbf{K}}_\mathrm{r}$ \cite{Rayleigh1877} except if specific conditions are met to allow for a Rayleigh damping model  \cite{CaugheyJAM1965,AdhikariJSV2006,AdhikariJVA2009}. Reference~\cite{HusseinPRB2009} has conveniently considered the Rayleigh damping scenario and demonstrated that dissipation, especially when intense, can generate unique phenomena in the band structure such as branch overtaking and wavenumber cut-offs and cut-ons. Additional analysis on the effects of damping on the band structure, including in generally damped models, is offered in Ref.~\cite{PhaniJVA2013}. \\
\indent Presently, Eq.~\eqref{eq:eom_mat_4} represents a nonlinear eigenvalue problem. In order to recover a linear form, we first apply a state-space transformation to Eq.~\eqref{eq:eom_mat_3} \cite{HusseinJAP2010}
\begin{equation}
\begin{bmatrix}
\mathbf{0} & \bar{\mathbf{M}}_\mathrm{r}\\
\bar{\mathbf{M}}_\mathrm{r} & \beta\bar{\mathbf{C}}_\mathrm{r}
\end{bmatrix}\dot{\bar{\mathbf{y}}}+
\begin{bmatrix}
-\bar{\mathbf{M}}_\mathrm{r} & \mathbf{0}\\
\mathbf{0} & \omega_0^2\bar{\mathbf{K}}_\mathrm{r}
\end{bmatrix}\bar{\mathbf{y}}=\mathbf{0},
\end{equation}
where $\bar{\mathbf{y}}^\mathrm{T}=[\dot{\bar{\mathbf{u}}}\;\;\bar{\mathbf{u}}]$. Assuming a state-space solution  $\mathbf{y}=\tilde{\mathbf{y}}\mathrm{e}^{\lambda{t}}$, where $\tilde{\mathbf{y}}$ is a complex wave amplitude vector, we formulate the following generalized linear eigenvalue problem in $\lambda$:
\begin{equation}\left(
\begin{bmatrix}
\mathbf{0} & \bar{\mathbf{M}}_\mathrm{r}\\
\bar{\mathbf{M}}_\mathrm{r} & \beta\bar{\mathbf{C}}_\mathrm{r}
\end{bmatrix}\lambda+
\begin{bmatrix}
-\bar{\mathbf{M}}_\mathrm{r} & \mathbf{0}\\
\mathbf{0} & \omega_0^2\bar{\mathbf{K}}_\mathrm{r}
\end{bmatrix}\right)\tilde{\mathbf{y}}=\mathbf{0}.
\end{equation}
The associated characteristic equation takes the form
\begin{equation}
\label{eq:char_eq_freq}
\lambda^4+a\lambda^3+b\lambda^2+c\lambda+d=0,
\end{equation}
where
\begin{subequations}
	\begin{align}
	&a=\frac{\beta[r_\mathrm{c}(1+r_\mathrm{m})+2r_\mathrm{m}(1-\cos\kappa{a})]}{r_\mathrm{c}},\\
	&b=\frac{r_\mathrm{c}r_\mathrm{k}\omega_0^2(1+r_\mathrm{m})+2r_\mathrm{m}(1-\cos\kappa{a})(r_\mathrm{k}\beta^2+r_\mathrm{c}\omega_0^2)}{r_\mathrm{c}r_\mathrm{k}},\\
	&c=\frac{2r_\mathrm{m}\beta\omega_0^2(r_\mathrm{c}+r_\mathrm{k})(1-\cos\kappa{a})}{r_\mathrm{c}r_\mathrm{k}},\\
	&d=\frac{2r_\mathrm{m}\omega_0^4(1-\cos\kappa{a})}{r_\mathrm{k}}.
	\end{align}
\end{subequations}
In general, the solutions $s=1,n$ (where $n$ is equal to the number of DOF) to Eq.~\eqref{eq:char_eq_freq} are complex and take the form
\begin{equation}
\label{eq:freq_cplx}
\lambda_s(\kappa)=-\xi_s(\kappa)\omega_{\mathrm{r},s}(\kappa)\pm\mathrm{i}\omega_{\mathrm{d},s}(\kappa).
\end{equation}
\noindent Specifically, $\omega_\mathrm{d}(\kappa)$ is the damped wave frequency and $\xi(\kappa)\omega_\mathrm{r}(\kappa)$ is the temporal rate of decay of the wave amplitude. The quantity $\xi(\kappa)$ is the dimensionless damping ratio (loss factor) and $\omega_\mathrm{r}(\kappa)$ is referred to as the ``resonant frequency". Explicitly, retaining the solutions for which $\mathrm{Im}[\lambda_s]\geq0$, the damped frequency relation for branch $s$ is
\begin{equation}
\label{eq:damped_freq}
\omega_{\mathrm{d},s}(\kappa)=\mathrm{Im}[\lambda_s(\kappa)],
\end{equation}
and the complementary wavenumber-dependent damping ratio relation is
\begin{equation}
\label{eq:damping_ratio}
\xi_s(\kappa)=-\frac{\mathrm{Re}[\lambda_s(\kappa)]}{|\lambda_s(\kappa)|}.
\end{equation}

%The proceeding results of the linear eigenvalue problem are valid only for real-valued wavenumbers, $\kappa=\kappa_\mathrm{R}$.  Although the quadratic eigenvalue methodology (discussed next) rectifies this apparent shortcoming and yields the complete propagating and evanescent band structure, it is not without the support of insights garnered from linear eigenvalue problem. Furthermore, the proceeding formulations are invaluable in the calculation of the effective mechanical properties covered in Part II, the companion to this work.

\subsection{Wavenumber Solutions from Quadratic Eigenvalue Problem} \label{sec:solution_wave}
The preceding linear eigenvalue formulation is applicable only for prescribed, real-valued wavenumbers, $\kappa=\kappa_\mathrm{R}$. In general, however, the wavenumber is complex regardless if we are considering driven waves or free waves, as explained in Section~\ref{sec:introd}. The wavenumber in complex form is expressed as $\kappa=\kappa_\mathrm{R}+\mathrm{i}\kappa_\mathrm{I}$, with $\kappa_\mathrm{R}$ representing the wave spatial oscillation and $\kappa_\mathrm{I}$ representing the spatial amplitude decay. In order to determine these quantities for a wave of a given frequency, $\lambda$, the problem must be reformulated to deliver wavenumber solutions in exchange.

In Eq. \eqref{eq:eom_mat_4}, we leave the value of the wavenumber (and, therefore, $\gamma$) to be determined. Taking the determinate of the coefficient matrix in Eq. \eqref{eq:eom_mat_4}, the characteristic equation in $\gamma$ takes the following quadratic form:
\begin{equation}
\label{eq:char_eq_wave}
\hat{A}\gamma^2+\hat{B}\gamma+1=0,
\end{equation}
where $\hat{A}=1$, $\hat{B}=\hat{B}_1/\hat{B}_2$ and 
\begin{subequations}
	\begin{align}
	\begin{split}
	\hat{B_1}&=\frac{\lambda^4}{r_\mathrm{m}}+\frac{(2+r_\mathrm{c})r_\mathrm{m}+r_\mathrm{c}}{r_\mathrm{m}r_\mathrm{c}}\beta\lambda^3+\left(2\frac{\beta^2}{r_\mathrm{c}}+\frac{(2+r_\mathrm{k})r_\mathrm{m}+r_\mathrm{k}}{r_\mathrm{m}r_\mathrm{k}}\omega_0^2\right)\lambda^2\\
	&\quad+\frac{2(r_\mathrm{c}+r_\mathrm{k})\beta\omega_0^2}{r_\mathrm{c}r_\mathrm{k}}\lambda+2\frac{\omega_0^4}{r_\mathrm{k}}, 
	\end{split} \\
	\hat{B_2}&=-\frac{[\lambda(\beta+\lambda)+\omega_0^2](r_\mathrm{k}\beta\lambda+r_\mathrm{c}\omega_0^2)}{r_\mathrm{c}r_\mathrm{k}}.
	\end{align}
\end{subequations}
Upon solving for $\gamma$, $\kappa_\mathrm{R}$ and $\kappa_\mathrm{I}$ are extracted as follows:
\begin{subequations}
	\label{eq:wavenumber}
	\begin{align}
	\kappa_\mathrm{R}=\frac{1}{a}|\mathrm{Re}[\mathrm{i\cdot ln}\gamma]|,\\
	\kappa_\mathrm{I}=\frac{1}{a}|\mathrm{Im}[\mathrm{i\cdot ln}\gamma]|.
	\end{align}
\end{subequations}

In the absence of energy dissipation, $\lambda=\mathrm{i}\omega$, where $\omega$ is a real number representing the wave frequency, and $\kappa$ is obtained directly by solving Eq.~\eqref{eq:char_eq_wave} for a given value of $\omega$. In the presence of energy dissipation, $\kappa$ is obtained also directly by solving Eq.~\eqref{eq:char_eq_wave} for waves with a prescribed frequency, $\omega$, i.e., driven waves. However, for free waves, energy dissipation results in the frequency being complex, as is the wavenumber, as discussed above. Thus the frequencies take the form  $\lambda=-\xi\omega_\mathrm{r}+\mathrm{i}\omega_\mathrm{d}$.

% Specifically, $\omega_\mathrm{d}$ is the damped wave frequency and $\lambda_\mathrm{R}=\xi\omega_\mathrm{r}$ is the rate of temporal decay of the wave amplitude where $\xi$ is the dimensionless damping ratio and $\omega_\mathrm{r}$ is referred to as the ``resonant frequency".

\begin{figure}[h]
	\centering
	\includegraphics[scale= 0.8]{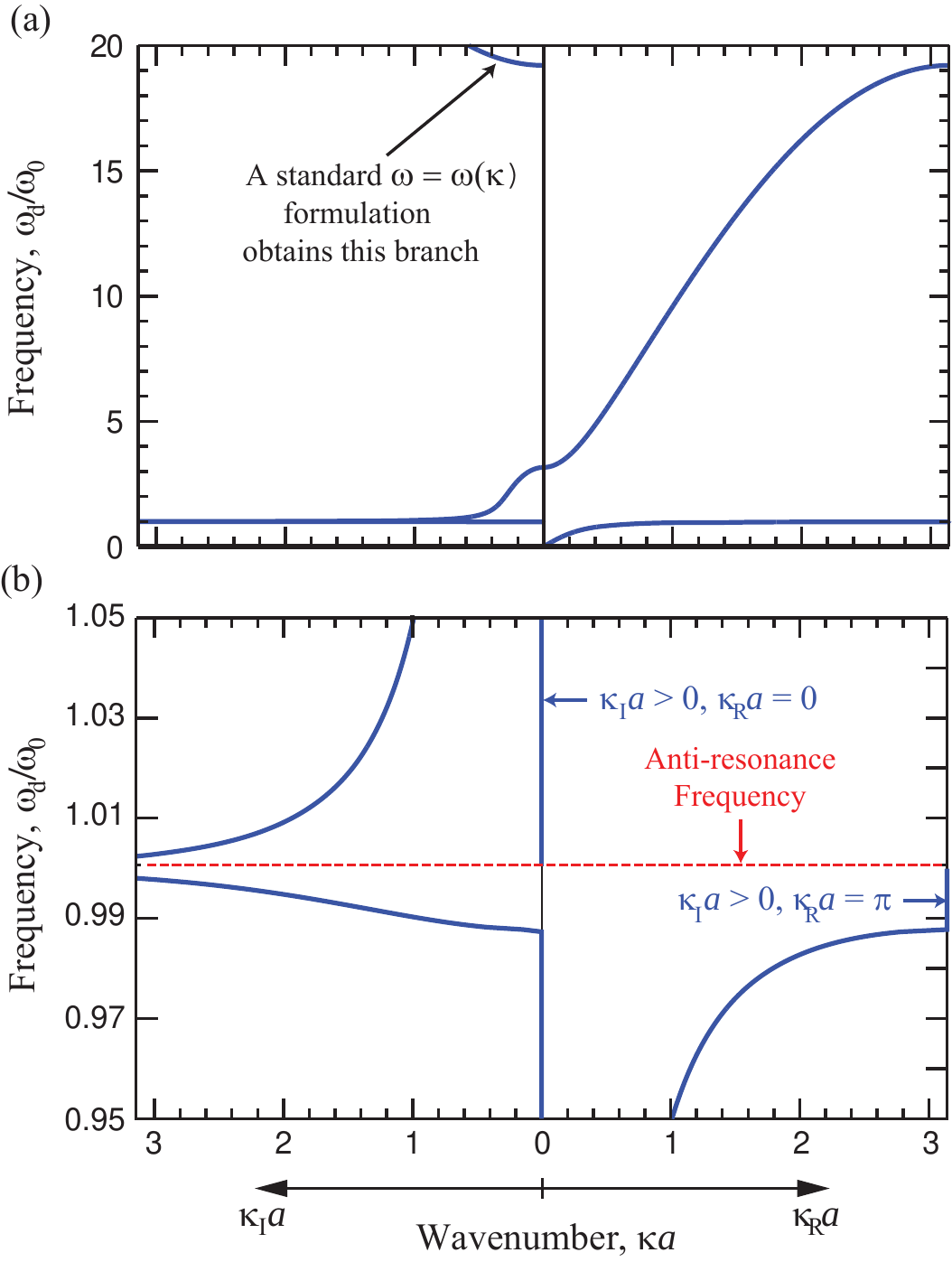}
	\caption{Undamped band structure of the locally resonant metamaterial of Fig.~\ref{fig:FIG_01_Model}a. (a) Uppermost branch is obtained by a standard $\omega=\omega(\kappa)$ formulation as well as by the proposed algorithm, thus providing a verification on the accuracy of the algorithm. (b) Illustration of wavenumber subsegments inside the band gap. Separated by an anti-resonance frequency, the lower subsegment has $\kappa_\mathrm{R}=\pi/a$ and the upper subsegment has $\kappa_\mathrm{R}=0$.}
	\label{fig:FIG_02_Resonance_Gap}
\end{figure}

From earlier work on viscous problems, e.g.,~\cite{HusseinJAP2010}, we understand that damped, free wave propagation produces two band diagrams representing the real and imaginary components of the frequency, respectively. Additionally, for locally resonant metamaterials, each free waves band diagram can be divided into band segments featuring only propagating waves (i.e., purely real wavenumbers) and band segments featuring evanescent waves. In the latter case, there are two possible subsegments: one with complex wavenumbers and one with purely imaginary wavenumbers, both of which take place only within a band gap. In a subsegment where the wavenumber is complex, the real part is equal to $\pi/a$. Thus a band gap bounded by pass bands from the bottom and the top has a subsegment where $\kappa_\mathrm{R}=0$ and a subsegment where $\kappa_\mathrm{R}=\pi/a$, as illustrated in Fig.~\ref{fig:FIG_02_Resonance_Gap}b. Following the state-space $\omega=\omega(\kappa)$ formulation in Sec.~\ref{sec:solution_freq}, the propagating segments are readily obtained~\cite{HusseinJAP2010}. However, a complete solution featuring all available imaginary wavenumbers is not possible with this formulation.

 % To determine the complete band structure for complex frequencies, including portions featuring complex wavenumbers, we turn to Eq. \eqref{eq:char_eq_wave}.

We approach this problem also by using Eq.~\eqref{eq:char_eq_wave}. However, a unique combination of the real and imaginary components of $\lambda$ does not exist without constraints. Following from the above discussion, a free wave does not allow any propagating component (other than $\kappa_\mathrm{R}=\pi/a$) to exist inside a band gap. Thus a damped, free wave at a particular frequency is either outside a band gap and is purely propagating or inside  a band gap and is evanescent. Based on these characteristics, we develop an algorithm to find the ($\xi,\omega_\mathrm{d}$) pair that satisfies Eq.~\eqref{eq:char_eq_wave} with $\kappa_\mathrm{I}>0$ which gives us the attenuation constant inside band gaps. The propagating portion of the band structure on the other hand is easily determined from Eqs.~\eqref{eq:damped_freq} and \eqref{eq:damping_ratio}. Nevertheless, the algorithm may be readily modified to determine the propagating part of the band structure as well. Although this procedure is developed following a quadratic eigenvalue formulation, a similar algorithm based on a linear formulation is a subject for future research.
\\\\
\textbf{Algorithm for all-complex band structure for free waves:}\\

\it{This algorithm is specific to the evanescent part of the band structure}.\rm
\begin{enumerate}
	\item Set $\beta>0$.
	\item Determine $\lambda_\mathrm{I}(\kappa)=\mathrm{Im}[\lambda(\kappa)]$ and $\lambda_\mathrm{R}(\kappa)=-\mathrm{Re}[\lambda(\kappa)]$ following the method in Sec. \ref{sec:solution_freq} for $\kappa=\kappa_\mathrm{R}$.
	%\item Set a target $\kappa_\mathrm{I}>0$ and a target $\kappa_\mathrm{R}=0,\pi$.
	\item Define a two-dimensional search space $\lambda_\mathrm{I}\textrm{-}\lambda_\mathrm{R}$ corresponding to the gap regions in the $\lambda_\mathrm{I}(\kappa)$ plot.
	\item Discretize the $\lambda_\mathrm{I}\textrm{-}\lambda_\mathrm{R}$ domain into a grid of points ($\lambda_{\mathrm{I},i},\lambda_{\mathrm{R},j}$).
	\item For each $\lambda_{i,j}=-\lambda_{\mathrm{R},j}+\mathrm{i}\lambda_{\mathrm{I},i}$, calculate $\kappa_{\mathrm{I},i,j}$ following the method in Sec. \ref{sec:solution_wave}.
	%\item For each $\gamma_{i,j}$, calculate $\kappa_{\mathrm{I},i,j}$
	\item Set a target value for $\kappa_\mathrm{I}$ (maintaining that $\kappa_\mathrm{I}>0$) and a target $\kappa_\mathrm{R}=0,\pi$.
	\item Of the $\lambda_{i,j}$ that produce $\kappa_{\mathrm{I},i,j}$ within tolerance of the target $\kappa_\mathrm{I}$, retain the one or more $\lambda_{i,j}$ that correspond to $\kappa_{\mathrm{R},i,j}$ closest to the target $\kappa_\mathrm{R}$.
	\item Extract $\omega_{\mathrm{d},{i,j}}$ and $\xi_{i,j}$ from the retained $\lambda_{i,j}$ using Eqs. \eqref{eq:damped_freq} and \eqref{eq:damping_ratio}, respectively.
	\item Repeat steps 6--8 for different target $\kappa_\mathrm{I}>0$ until the evanescent band structure is constructed.\\
\end{enumerate}

The outcome of the above algorithm is an approximation of the evanescent frequencies and the damping ratios for a particular damping intensity. In addition to setting up a finer grid over the $\lambda_\mathrm{R}\textrm{-}\omega_\mathrm{d}$ domain for a better approximation of $\lambda_{i,j}$, a more clear picture of the evanescent band structure obviously results from decreasing the separation between subsequent target $\kappa_\mathrm{I}$ values.
\begin{figure}[h]
	\centering
	\includegraphics[scale= 0.8]{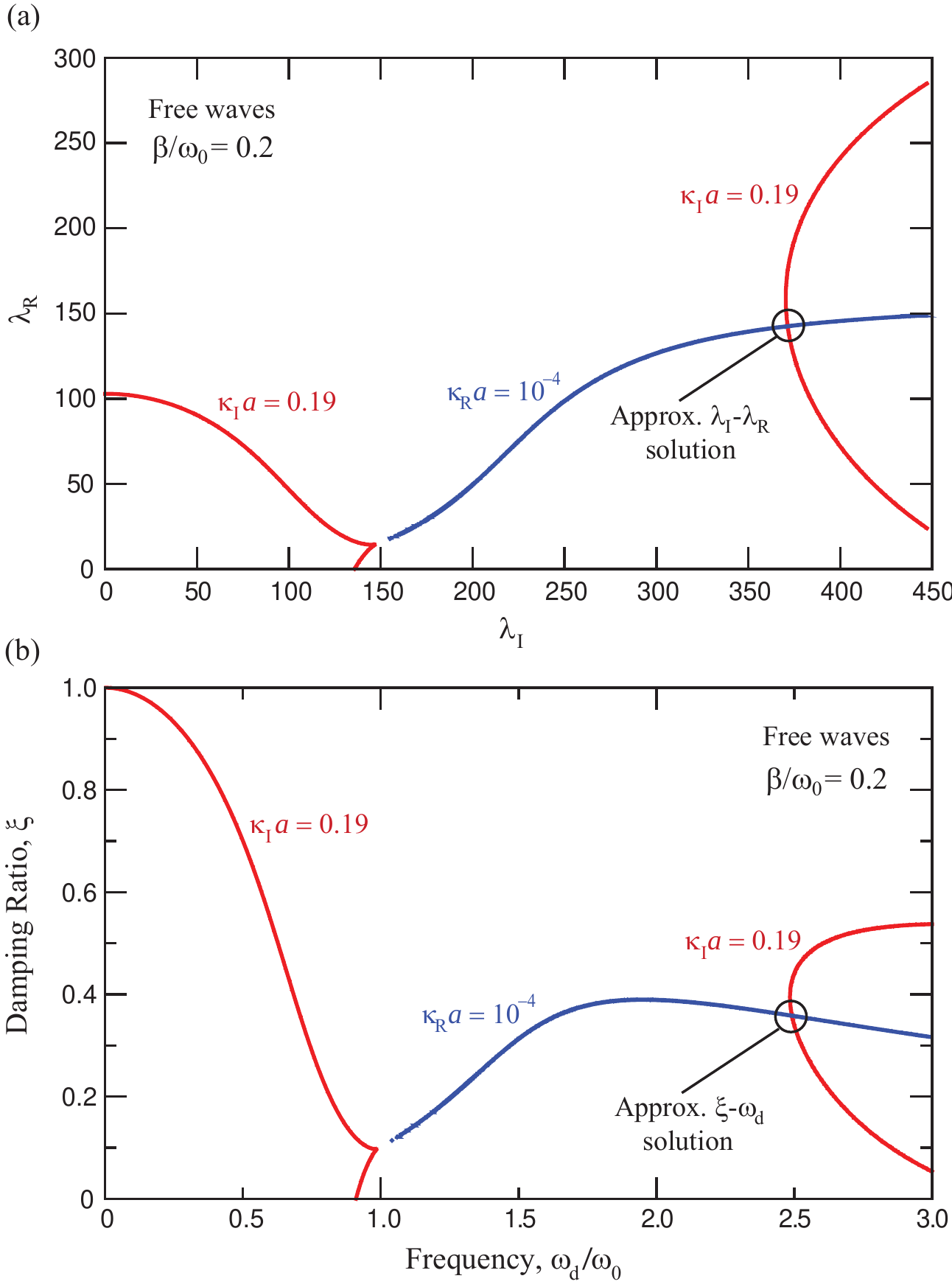}
	\caption{Demonstration of all-complex band-structure calculation algorithm. (a) For a $\beta\neq0$ value, a $\lambda_\mathrm{I}\text{-}\lambda_\mathrm{R}$ search space is defined corresponding to the gap regions seen in $\lambda_\mathrm{I}$ following Eq. \eqref{eq:freq_cplx} (Steps 1--3). After discretizing the 2D search space, Eq. \eqref{eq:wavenumber} is used to determine the set of $\lambda_{i,j}$ giving approximations (within a preset set tolerance threshold) to the target $\kappa_\mathrm{I}a=0.19$ (red) (Steps 4-6). Of these, the one $\lambda_{i,j}$ that best approximates the target $\kappa_\mathrm{R}a\approx0$ (blue) is taken as the solution from which $\omega_{\mathrm{d},{i,j}}$ and $\xi_{i,j}$ are extracted (Steps 7 and 8). (b) Demonstration of the same procedure in the corresponding $\xi\textrm{-}\omega_\mathrm{d}$ domain.}
	\label{fig:FIG_03_Loops}
\end{figure}

Figure~\ref{fig:FIG_03_Loops} gives a visual example of an application of the algorithm for a specific set of material parameters ($r_\mathrm{m}=9$, $r_\mathrm{c}=1$, $r_\mathrm{k}=1/10$, $\omega_0=149.07$ rad/s and $\beta/\omega_0=0.2)$. Shown later in Figs.~\ref{fig:FIG_04_Dispersion}b and \ref{fig:FIG_04_Dispersion}c are the band-gap ranges used to bound the $\lambda_\mathrm{I}\text{-}\lambda_\mathrm{R}$ domain for this example. In the closed, discretized $\lambda_\mathrm{R}\textrm{-}\lambda_\mathrm{I}$ domain of Figure \ref{fig:FIG_03_Loops}a,
%[$\xi$ extracted from $\lambda_\mathrm{R}$ according to Eq.~\eqref{eq:damping_ratio}]
we isolate the set of points that satisfy, separately, the conditions (targets) $\kappa_\mathrm{I}a=0.19$ and $\kappa_\mathrm{R}a\approx0$. At the intersection of the $\kappa_\mathrm{I}a=0.19$ and $\kappa_\mathrm{R}a\approx0$ loci is the $\lambda_{i,j}$ value which is retained as an approximate solution. The intersection moves along the $\kappa_\mathrm{R}a\approx0$ curve for each new target $\kappa_\mathrm{I}a$, generating new $\lambda_{i,j}$ approximations which ultimately construct the evanescent frequency and damping ratio band diagrams over the range of the targeted $\kappa_{\mathrm{I}}$ values. More than one physical solution may exist for a given $\kappa_{\mathrm{I}}$. To validate the algorithm, we examine the $\kappa_\mathrm{I}$ values corresponding to  the uppermost band gap, which for an undamped system is practically unbounded (see Fig.~\ref{fig:FIG_02_Resonance_Gap}a). This upper branch is unique in that it may be obtained by both the state-space $\omega=\omega(\kappa)$ formulation given in Sec.~\ref{sec:solution_freq} (as demonstrated in Ref.~~\cite{HusseinJAP2010}) as well as by the algorithm. Both independent routes yield the same $\kappa_\mathrm{I}$ values for any level of damping intensity. 

\section{Theory: Equivalent Mass Model}
%To demonstrate the consequences of energy dissipation in acoustic metamaterials, the dispersion and the effective properties, we consider the nested 1D lumped parameter mass-spring-dashpot model in Fig. \ref{fig:FIG_01_Model}.
Figure \ref{fig:FIG_01_Model}b shows the equivalent lattice model, which we require to exhibit the same dynamic behavior as the original nested-mass model. The equivalent lattice model has the same construction as the original metamaterial (i.e., $k_1$ and $c_1$ are the same) except the motion of the internal resonator is considered unobservable, although its influence is accounted for by $m_\mathrm{e}$, the effective mass. The value $m_\mathrm{e}$ varies with frequency, satisfying its own dynamic equilibrium, and concurrently matching the complex frequency of the original mass-in-mass metamaterial. This concept was also applied in Ref.~\cite{MiltonPRSA2007} where a rigid bar conceal periodically distributed internal resonators. Expectedly, at low frequencies, the value of $m_\mathrm{e}$ converges to the static value, $m_\mathrm{st}=m_1+m_2$. The terms ``static" and ``nominal" are used to indicate values attained in the long-wavelength limit.

The equation of motion for each degree of freedom in the equivalent lattice model (Fig. \ref{fig:FIG_01_Model}b) is given by:
\begin{subequations}
	\label{eq:eom_nodal_equiv}
	\begin{align}
	&m_\mathrm{e}\ddot{u}_\mathrm{e}+c_1(\dot{u}_\mathrm{e}-\dot{u}_\mathrm{L})+k_1(u_\mathrm{e}-u_\mathrm{L})=f_\mathrm{e},\\
	&m_\mathrm{L}\ddot{u}_\mathrm{L}+c_1(\dot{u}_\mathrm{L}-\dot{u}_\mathrm{e})+k_1(u_\mathrm{L}-u_\mathrm{e})=f_\mathrm{L}.
	\end{align}
\end{subequations}
Assembling Eqs. \eqref{eq:eom_nodal_equiv} into a system of equations as in Eq. \eqref{eq:eom_mat_1}, the mass, damping, and stiffness matrices are defined as follows:
\begin{subequations}
	\begin{align}
	&\mathbf{M}=
	\begin{bmatrix}
	m_\mathrm{e} & 0\\
	0 & 0
	\end{bmatrix},\\
	&\mathbf{C}=\frac{c_2}{r_\mathrm{c}}
	\begin{bmatrix}
	1 & -1\\
	-1 & 1
	\end{bmatrix},\\
	&\mathbf{K}=\frac{k_2}{r_\mathrm{k}}
	\begin{bmatrix}
	1 & -1\\
	-1 & 1
	\end{bmatrix},
	\end{align}
\end{subequations}
with the nodal displacements and forces organized as $\mathbf{u}^\mathrm{T}=[u_\mathrm{e}\;\;u_\mathrm{L}]$ and $\mathbf{f}^\mathrm{T}=[f_\mathrm{e}\;\;f_\mathrm{L}]$, respectively. The essential set of displacements is simply $\bar{u}=u_\mathrm{e}$. Assuming a real-valued wavenumber, we apply the Bloch boundary conditions through $\mathbf{T}^\mathrm{T}=[1\;\;\gamma]$. This leads to
\begin{equation}
\label{eq:eom_equiv_1}
\frac{(1+r_\mathrm{m})m_\mathrm{r}}{r_\mathrm{m}}\ddot{u}_\mathrm{e}+2(1-\cos\kappa{a})\left(\frac{\beta}{r_\mathrm{c}}\dot{u}_\mathrm{e}+\frac{\omega_0^2}{r_\mathrm{k}}u_\mathrm{e}\right)=0,
\end{equation}
in which the effective mass ratio has been normalized by the static mass,  $m_\mathrm{r}=m_\mathrm{e}/m_\mathrm{st}$, and the previously defined ratios $r_\mathrm{m}$, $r_\mathrm{c}$, $r_\mathrm{k}$, $\beta$, and $\omega_0^2$ have been taken advantage of. Applying the harmonic displacement solution $u_\mathrm{e}=\tilde{u}_\mathrm{e}\mathrm{e}^{\lambda{t}}$, Eq. \eqref{eq:eom_equiv_1} gives
\begin{equation}
\label{eq:eom_equiv_2}
\frac{(1+r_\mathrm{m})m_\mathrm{r}}{r_\mathrm{m}}\lambda^2+2\left(\frac{\beta}{r_\mathrm{c}}\lambda+\frac{\omega_0^2}{r_\mathrm{k}}\right)(1-\cos\kappa{a})=0,
\end{equation}
however, in preparation for substitution, Eq. \eqref{eq:eom_equiv_2} is algebraically manipulated into the following form:
\begin{equation}
\label{eq:eom_equiv_3}
-\frac{(1+r_\mathrm{m})m_\mathrm{r}}{\left(\frac{\beta}{r_\mathrm{c}}\lambda+\frac{\omega_0^2}{r_\mathrm{k}}\right)r_\mathrm{m}}\lambda^2=2(1-\cos\kappa{a}).
\end{equation}

Now, we tie the dynamic characteristics of the equivalent model to the original metamaterial model by simultaneously substituting Eq. \eqref{eq:eom_equiv_3} into Eqs. \eqref{eq:mat_def_freq_b} and \eqref{eq:mat_def_freq_c}.
\begin{subequations}
	\begin{align}
	&\bar{\mathbf{C}}_\mathrm{r}=
	\begin{bmatrix}
	-\frac{(1+r_\mathrm{m})m_\mathrm{r}}{\left(\frac{\beta}{r_\mathrm{c}}\lambda+\frac{\omega_0^2}{r_\mathrm{k}}\right)r_\mathrm{m}r_\mathrm{c}}\lambda^2+1 & -1\\
	-1 & 1
	\end{bmatrix},\\
	&\bar{\mathbf{K}}_\mathrm{r}=
	\begin{bmatrix}
	-\frac{(1+r_\mathrm{m})m_\mathrm{r}}{\left(\frac{\beta}{r_\mathrm{c}}\lambda+\frac{\omega_0^2}{r_\mathrm{k}}\right)r_\mathrm{m}r_\mathrm{k}}\lambda^2+1 & -1\\
	-1 & 1
	\end{bmatrix}.
	\end{align}
\end{subequations}

Using these new matrix definitions for the damping and stiffness matrices, we proceed with Bloch state-space treatment. This process delivers the following characteristic equation:
\begin{equation}
[1-m_\mathrm{r}(1+r_\mathrm{m})]\lambda^4+(1-m_\mathrm{r})(1+r_\mathrm{m})(\beta\lambda+\omega_0^2)\lambda^2=0,
\end{equation}
which we subsequently solve for $m_\mathrm{r}(\lambda)$:
\begin{equation}
\label{eq:eff_mass}
m_\mathrm{r}(\lambda)=1-\frac{r_\mathrm{m}\lambda^2}{(1+r_\mathrm{m})[\lambda(\beta+\lambda)+\omega_0^2]}.
\end{equation}
%Notice the absence of any wavenumber dependence. As mentioned previously, the value of the effective mass must only satisfy the equilibrium equation of the equivalent model while matching the frequency of the metamaterial.
The quantity $m_\mathrm{r}(\lambda)$ is evaluated for the damped free waves problem by substituting $\lambda$ with the values obtained from the solution of Eq.~\eqref{eq:char_eq_freq} and the execution of the algorithm.

\section{Numerical Examples}
In this section, we present a suite of numerical examples to demonstrate all-complex band structures for damped free wave propagation. The examples are for the mass-in-mass metamaterial model shown in Fig.~\ref{fig:FIG_01_Model}a, which allows us to examine, in a general manner, the role of dissipation in acoustic/elastic metamaterials, both in terms of the dispersion curves and the effective mass. The damping intensity is varied to give a broad representation of dissipative effects. The same set of material parameters used for the Fig.~\ref{fig:FIG_03_Loops} demonstration is used here as an example for our exposition. These parameters are consistent with the relatively high material contrast characteristic of metamaterials and are similar to those selected in Ref.~\cite{HusseinJAP2010} thus providing an opportunity for direct comparison with some of the results presented in that publication.
\begin{figure}[h]
	\centering
	\includegraphics[scale= 0.8]{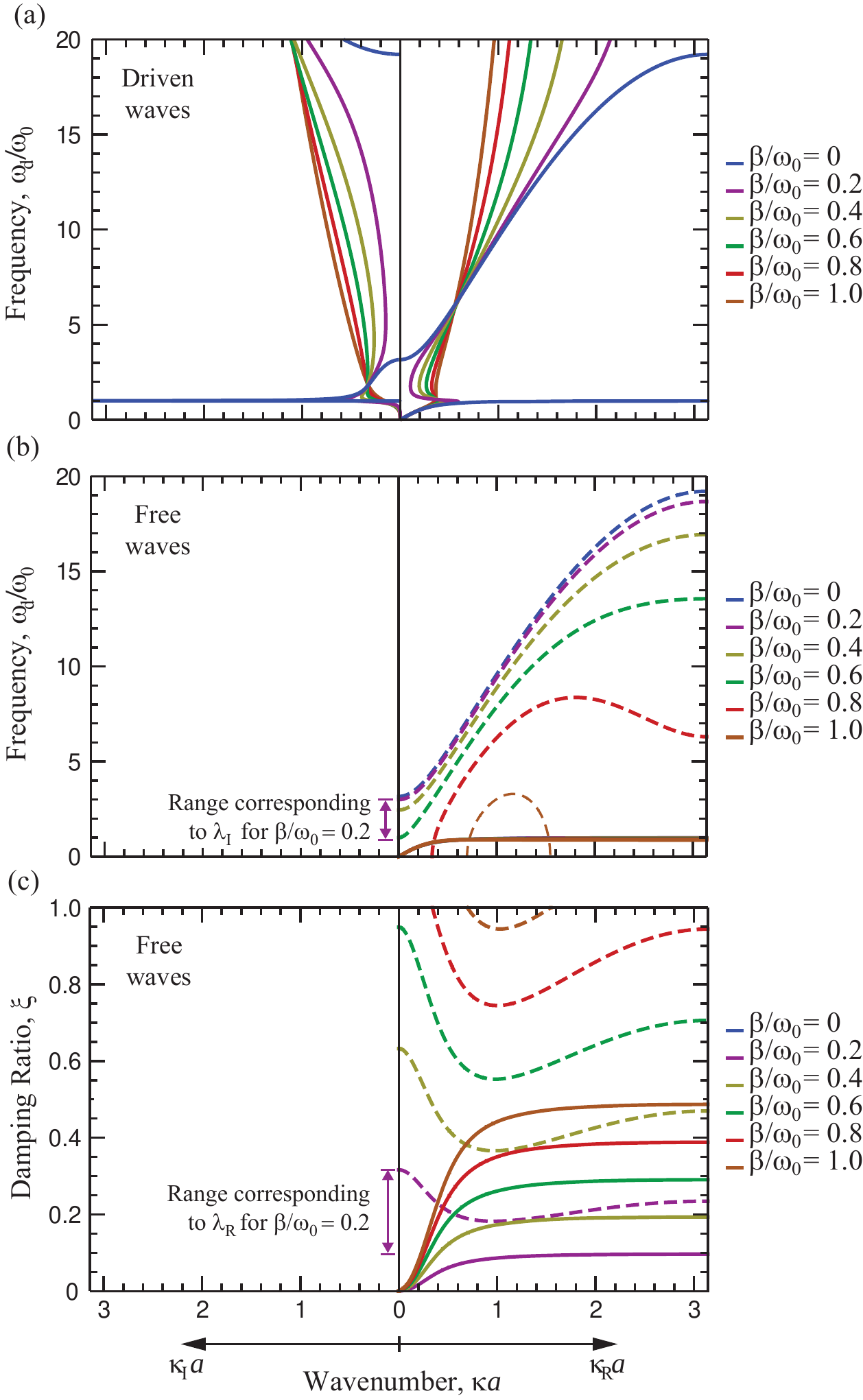}
	\caption{Dispersion band diagrams. (a) The frequency band diagram for driven waves generated by direct solving of $\kappa=\kappa(\lambda)$ where, for $\beta/\omega_0>0$, $\kappa$ is complex for all prescribed $\lambda=\mathrm{i}\omega$. The incomplete frequency (b) and damping ratio (c) band diagrams for free waves generated by direct solving of $\lambda=\lambda(\kappa)$ for all prescribed $\kappa=\kappa_\mathrm{R}$ where $\kappa_\mathrm{R}\in[0,\pi/a]$.}
	\label{fig:FIG_04_Dispersion}
\end{figure}

As a consequence of the unit cell's two-DOF character, in the undamped case ($\beta/\omega_0=0$), there are two modes of wave propagation comprising an acoustical (lower) branch and an optical (upper) branch. These are shown in the dispersion band diagrams in Fig.~\ref{fig:FIG_04_Dispersion}. Separating the two branches, there is a resonance-induced band-gap region where $\kappa_\mathrm{R}=0,\pi$ and $\kappa_\mathrm{I}>0$ describing evanescent modes for which waves do not propagate and their amplitudes spatially decay at rates dictated by the value of $\kappa_\mathrm{I}$. In addition to a closed band-gap region, an unbounded region of evanescent modes exists above the optical branch. For compact presentation, we plot the imaginary wavenumbers within the range $\kappa_\mathrm{I}a\in[0,\pi]$. Without an algorithm, a band diagram with complex wavenumbers can only be determined for damped waves with prescribed real frequencies, that is, $\kappa=\kappa(\lambda=0+\mathrm{i}\omega)$ (Fig.~\ref{fig:FIG_04_Dispersion}a). In this scenario, the band gap closes and becomes progressively more obscure with greater damping intensity. Alternatively, the complex frequencies of damped free waves can only be directly calculated for $\lambda=\lambda(\kappa=\kappa_\mathrm{R}+\mathrm{i}0)$ (Fig.~\ref{fig:FIG_04_Dispersion}b). The frequency band gaps for free waves are shown to shift but remain intact with increased damping intensity. At extreme damping levels, however, branch overtakings and cut-offs and/or cut-ins are observed to take place. Figure \ref{fig:FIG_04_Dispersion} contrasts each of these ``either/or" scenarios for various values of $\beta/\omega_0$. A complete band structure description featuring generally a complex $\lambda$ and a complex $\kappa$ is made possible by the algorithm presented in Section~\ref{sec:solution_wave}.
\begin{figure}[h]
	\centering
	\includegraphics[scale= 0.8]{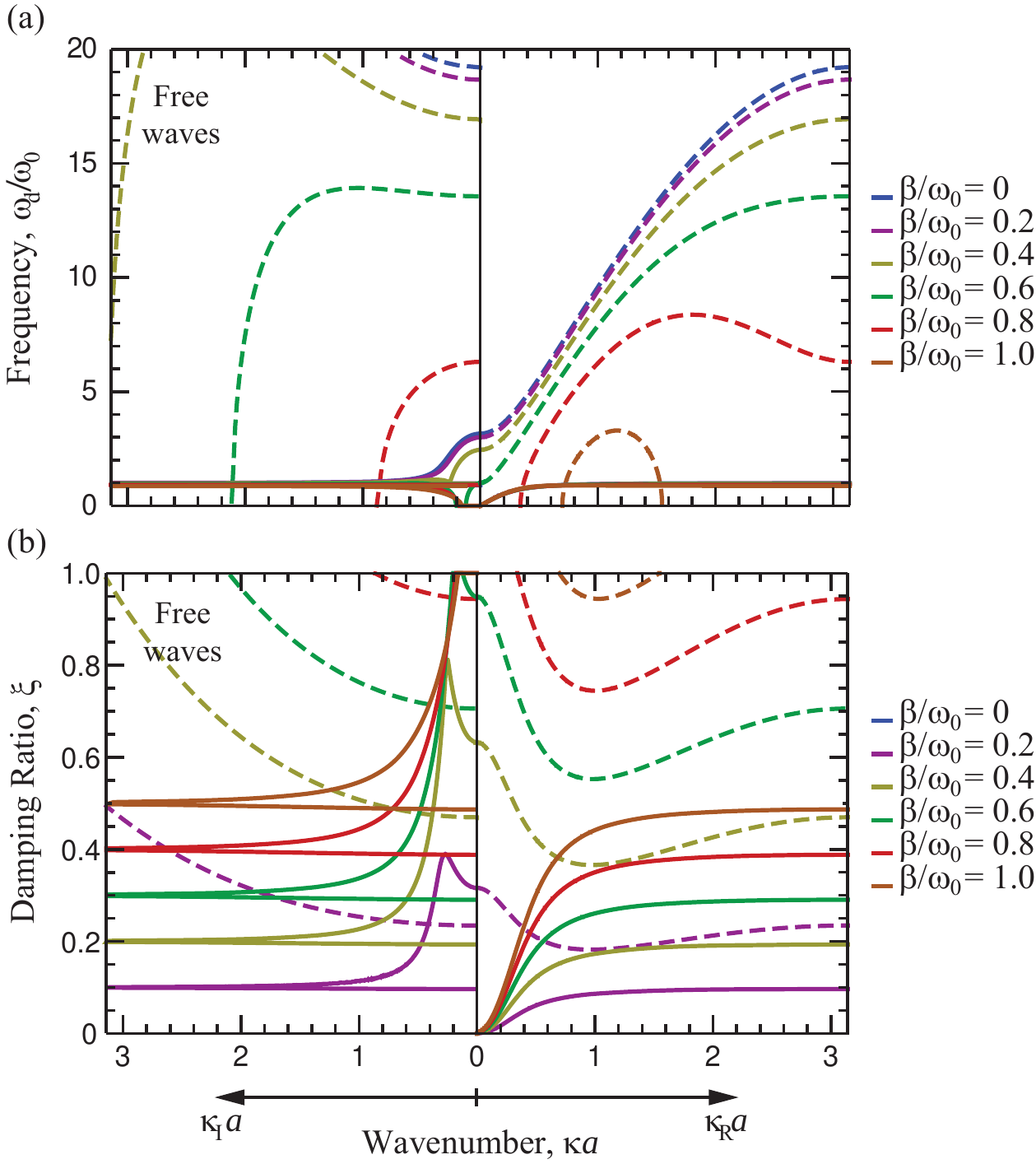}
	\caption{All-complex dispersion curves. The complete frequency (a) and damping ratio (b) band diagrams for free waves generated by the proposed algorithm for all prescribed $\kappa=\kappa_\mathrm{R}+\mathrm{i}\kappa_\mathrm{I}$ where $\kappa_\mathrm{R},\kappa_\mathrm{I}\in[0,\pi/a]$.}
	\label{fig:FIG_05_Dispersion}
\end{figure}

Now, we apply the algorithm and present the resulting all-complex band structure in Fig.~\ref{fig:FIG_05_Dispersion}. This band structure is, by definition, for the damped free waves case. Inspection of the effects of damping can now be done for both the propagating and evanescent modes. We note an intense responsiveness from the optical branch and the upper portion of the evanescent curves to increases in dissipation, in contrast to the still significant but milder effects on the acoustical branch and the lower portion of the evanescent curves. As damping increases, the concavity of the optical branch in Fig.~\ref{fig:FIG_05_Dispersion}a changes. This points to a damping-induced change in the sign of the wave group velocity from positive (longer wavelengths) to negative (shorter wavelengths). In addition, as mentioned above, the more rapid decent of the optical branch compared to the acoustical branch closes the band gap and inevitably leads to branch overtaking. Thus the optical and acoustical modes exchange order in the frequency spectrum over specific wavenumber values. Similarly in the evanescent modes, damping will induce a change in concavity in the upper portion of the $\kappa_\mathrm{I}$ curves. One interesting aspect of the behavior displayed in Fig.~\ref{fig:FIG_05_Dispersion}b is that the damping ratio of the lower, non-propagating portion of the evanescent curves appears to peak at approximately $\kappa_\mathrm{I}a\approx0.25$ regardless of the level of damping.

\begin{figure}[h]
	\centering
	\includegraphics[scale= 0.8]{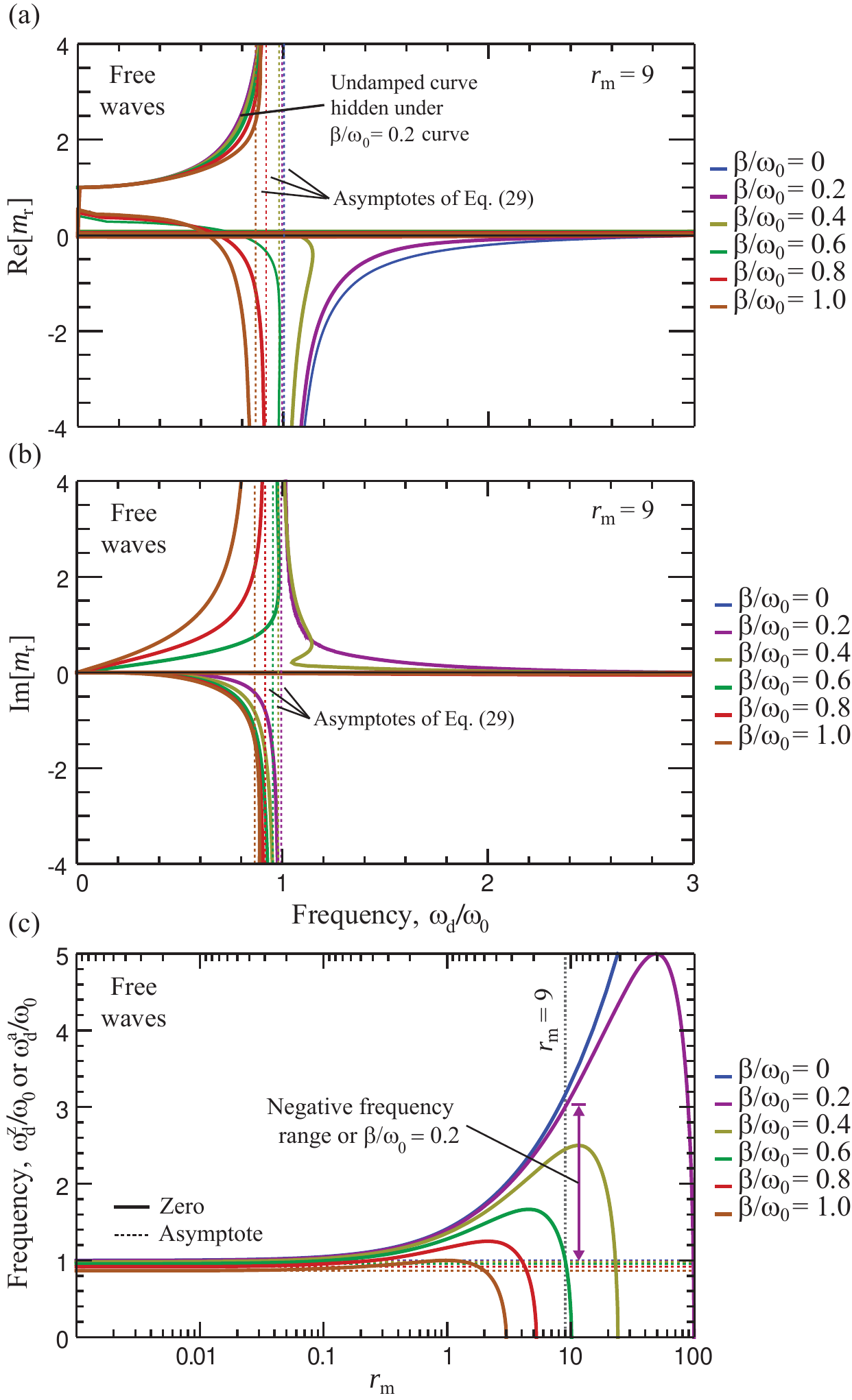}
	\caption{Dynamic effective mass curves. (a) Real component of effective mass; (b) imaginary component of effective mass; (c) zero (solid curves) and asymptote (dashed curves) of Eq.~\eqref{eq:eff_mass} enclosing the region of negative effective mass. The dashed vertical lines in (a) and (b) represent the real and imaginary asymptotes of Eq.~\eqref{eq:eff_mass}, respectively. In (c), the negative frequency range for $\beta/\omega_{0}=0.2$ is highlighted as an example.}
	\label{fig:FIG_06_Effective_Mass}
\end{figure}

We turn to the effective mass, which is shown by Eq.~\eqref{eq:eff_mass} to be frequency dependent. Without damping, the effective mass approaches its static value in the long wavelength limit ($\kappa,\lambda\rightarrow0$), as expected. That is, $m_\mathrm{r}\rightarrow1$ in the long wavelength limit, indicating that $m_\mathrm{e}\rightarrow{m_\mathrm{st}}$. Several studies have shown the effective mass to become negative over a frequency range in the region of the band gap, e.g., Refs.~\cite{HuangIJES2009,NematNasserAIP2011}. Here and in Refs.~\cite{HusseinPRB2009,HusseinJAP2010}, we see that viscous damping narrows and eventually closes the band gap. Thus the natural question is: how does damping affect the effective mass? This question is addressed by using Eq.~\eqref{eq:eff_mass} with the complex frequencies determined by the algorithm in Sec.~\ref{sec:solution_wave}. Figure~\ref{fig:FIG_06_Effective_Mass} shows that damping acts to narrow the frequency region at which the effective mass is negative. Also shown is that in the presence of damping, increasing $r_\mathrm{m}$ widens this frequency region up to a point after which the effect abruptly reverses and the region narrows.

%\textcolor{red}{\st{First, we discuss further the effective mass in the absence of damping. In this case, the effective mass is negative near the resonance frequency of the internal resonator and extending upwards in frequency (this negativity, however, is only over a relatively narrow band). The value of the effective mass, on the other hand, is positive at frequencies lower than the upper boundary of the frequency band gap.}}

We observe that damping causes the effective mass to separate into real and imaginary components, as presented in Figs.~\ref{fig:FIG_06_Effective_Mass}a and ~\ref{fig:FIG_06_Effective_Mass}b, respectively, and still displays negativity over certain frequency ranges. In Fig.~\ref{fig:FIG_06_Effective_Mass}a, regardless of the damping intensity, the value of the real component of the effective mass converges to that of the static mass as $\lambda\rightarrow0$.  As expected, the imaginary component of the effective mass (Fig. \ref{fig:FIG_06_Effective_Mass}b) is zero under the same condition. In Figs.~\ref{fig:FIG_06_Effective_Mass}a and ~\ref{fig:FIG_06_Effective_Mass}b, the dashed vertical lines are the asymptotes of, respectively, the real and imaginary components of the normalized effective mass obtained by setting the denominator of Eq.~\eqref{eq:eff_mass} equal to zero and solving for the frequency for the chosen value of $r_\mathrm{m}$. In the undamped case, this marks the bifurcation $\kappa_\mathrm{R}a$ experiences as the wave frequency crosses the resonance frequency from below (in the band gap, $\kappa_\mathrm{R}a=\pi$ switches to $\kappa_\mathrm{R}a=0$), but such a correlation cannot be made in the presence of damping. As seen in Fig. \ref{fig:FIG_05_Dispersion}, overdamping causes portions of the band structure to collapse to $\omega_\mathrm{d}/\omega_{0}=0$ (even far from the long wavelength limit). Consequently, in the frequency dependent $m_\mathrm{r}$ diagram of Fig. \ref{fig:FIG_06_Effective_Mass}a, $m_\mathrm{e}$ converges to ${m_\mathrm{st}}$ at one instance of $\omega_\mathrm{d}=0$ (long wavelength limit) but tends toward another value at a separate instance of $\omega_\mathrm{d}/\omega_{0}=0$ (effect of damping). We also observe that at high damping, there are no longer any frequencies in which the effective mass is only negative. Although the closure of the frequency region of negative effective mass may be difficult to distinguish for some damping scenarios in Fig. \ref{fig:FIG_06_Effective_Mass}a, the effect is made clear in Fig. \ref{fig:FIG_06_Effective_Mass}c.

The metamaterial considered in our model relies on a massive internal resonator to produce a negative effective mass. In Fig.~\ref{fig:FIG_06_Effective_Mass}c, we illustrate the importance of the internal resonator by varying $r_\mathrm{m}$, which has the effect of changing the mass of $m_1$ while keeping all other material parameters constant. The solid curves in Fig.~\ref{fig:FIG_06_Effective_Mass}c are produced by setting Eq.~\eqref{eq:eff_mass} equal to zero and solving for the complex $\lambda(r_\mathrm{m})$. The frequency at which the effective mass becomes zero marks the frequency at which there is a sign change in the value of the effective mass for both the real and imaginary components. The dashed curves in Fig.~\ref{fig:FIG_06_Effective_Mass}c represent the asymptotes and are the result of setting the denominator of Eq.~\eqref{eq:eff_mass} equal to zero and solving for the complex $\lambda(r_\mathrm{m})$. These two curves bound the frequency ranges where $\mathrm{Re}[m_\mathrm{r}]<0$. When $r_\mathrm{m}\ll1$, that is, when $m_1\gg{m_2}$, the effect of the internal resonator diminishes and the identity of our metamaterial becomes ambiguous as it approaches the dynamic behavior of an ordinary phononic crystal. Consequently, in Fig.~\ref{fig:FIG_06_Effective_Mass}c, as $m_2$ loses influence, the frequency region over which the effective mass is negative shrinks to near nonexistence.

\section{Conclusions}
In this work, we investigated the notion that damped free waves are in principle governed by a dispersion relation in which both the frequency and the wavenumber are generally complex, not only one or the other being complex as commonly assumed. An algorithm based on a $\kappa=\kappa(\lambda)$ formulation, guided by first solving the $\lambda=\lambda(\kappa=\kappa_\mathrm{R}+\mathrm{i}0)$ problem, was presented for 1D periodic chains. The algorithm was applied to a mass-in-mass unit cell representing a viscously damped locally resonant acoustic/elastic metamaterial. This analysis allows one to examine the effects of damping not only on the propagating modes of free waves, but also on the evanescent modes. For both mode sets, the effects of damping appear in both the frequency and the damping ratio band diagrams. A dynamic effective mass for the damped metamaterial model was also calculated and was shown to exhibit negative values over a frequency region near the band gap, as in the undamped case. However, for relatively high levels of damping, no frequencies are found in which the effective mass is only negative. Future work will explore multi-dimensional systems and the effects of other types of damping, e.g., nonviscous damping, in the context of the generalized, all-complex Bloch formulation presented in this paper. 

\section{Acknowledgment}
This research has been supported by the National Science Foundation Graduate Research Fellowship Grant No. DGE 1144083 and CAREER Grant No. 1254931. Support was also provided by the Department of Education GAANN program.

\bibliographystyle{ieeetr}
\bibliography{./References}
\end{document}